\documentclass[12pt]{article}

\usepackage{bbold}
\usepackage{amssymb}
\usepackage{graphics}
\usepackage{epsfig}

\newcommand{\Vec}[1]{\vec{#1}\hspace{.7mm}}

\renewcommand{\ss}{\scriptstyle}

\def\12{\frac{1}{2}}

\def\ch{{\rm cosh}\hspace{.5mm}}
\def\sh{{\rm sinh}\hspace{.5mm}}

\def\sss{\scriptscriptstyle}

\def\D{{\cal D}}

\def\wt#1{\widetilde{#1}}

\def\sss{\scriptscriptstyle}

\def\d{\partial}
\def\m{\mu}
\def\n{\nu}

\def\e{\epsilon}

\def\be{\begin{equation}}
\def\ee{\end{equation}}
\def\beq{\begin{equation}}
\def\eeq{\end{equation}}
\def\bea{\begin{eqnarray}}
\def\eea{\end{eqnarray}} 
\def\beqa{\begin{equation}\begin{array}{l}}
\def\eeqa{\end{array}\end{equation}}

\def\eqn#1{(\ref{#1})}

\def\eqref#1{eq.~(\ref{eq:#1})}

\def\nn{\nonumber}

\begin{document}

\thispagestyle{empty}
\begin{flushright}
\framebox{\small BRX-TH~511
}\\
\end{flushright}

\vspace{.8cm}
\setcounter{footnote}{0}
\begin{center}
{\Large{\bf
Arbitrary Spin Representations in de Sitter from 
dS/CFT with Applications to dS~Supergravity 
    }\\[10mm]}

{\sc S. Deser$^\sharp$
and A. Waldron$^\flat$
\\[6mm]}

{\em\small  
${}^\sharp$Physics Department, Brandeis University,\\ 
Waltham, MA 02454, 
USA\\ {\tt deser@brandeis.edu}}\\[5mm]

{\em\small ${}^\flat$Department of Mathematics, University
of California,\\
Davis, CA 95616, USA\\
{\tt wally@math.ucdavis.edu}}\\[5mm]

{\small \today}\\[1cm]

\bigskip

\bigskip

{\sc Abstract}\\

\end{center}

{\small
\begin{quote}

We present a simple group representation analysis of massive, and 
particularly ``partially massless'', fields of arbitrary 
spin in de Sitter spaces 
of any dimension. 
The method uses bulk to boundary propagators to relate 
these fields to Euclidean conformal ones at one dimension lower.
These results are then used to revisit an old question:
can a consistent de Sitter supergravity be 
constructed, at least within its intrinsic horizon?

\end{quote}

\newpage





\section{Introduction}

In addition to (perhaps) being our macroscopic habitat~\cite{Carroll:2000fy}, 
the conformally flat, constant curvature, de Sitter (dS) 
space has 
long been a 
seminal laboratory for field theory in curved backgrounds, being the 
simplest possible generalization of the Minkowski vacuum.
Indeed,~dS effects on a spin 1/2 field's 
propagation were first studied seven decades ago~\cite{Dirac:1935}.
Ever since, systems 
of all spins and masses have been analyzed in both~dS and anti de Sitter
(AdS), with obvious connections to string expansions\footnote{A
detailed analysis of partially massless representations has 
previously been given
by solving the Laplace equation on spheres for higher spin harmonics
and then analytically continuing these solutions to~dS
space~\cite{Higuchi:1987wu}. Although this analysis has the advantage 
of applying to the entire~dS space, it is rather complicated
and the simple interpretation of partially massless fields in terms
of missing lower helicity states is obscured.}.
A particularly striking effect, that of partial masslessness or partial gauge 
invariance, arising first at spin 2~\cite{Deser:1983tm,Deser:1984mm}, 
was recently discussed by us in some 
detail~\cite{Deser:2001xr,Deser:2001wx,Deser:2001us,Deser:2001pe,Deser:2001dt}. 
(see also~\cite{Zinoviev:2001dt,Dolan:2001ih}).
The purpose of the present work is 
twofold :  First, we formalize the kinematics underlying these 
properties in a group theoretic context, using a~dS/CFT
correspondence between $\rm dS_{n+1}$ 
and Euclidean space ${\mathbb R}^n$. This will 
exhibit the details of partial masslessness for all spin, in all 
dimensions $d=n+1$, as 
well as the ensuing propagation properties at these 
thresholds\footnote{Note that
the partial masslessness phenomenon occurs in AdS as well. In that
case, however, only strictly massless representations are unitary. The
representation theory of these fields is much easier to understand,
since the algebra can be graded with respect to the generator
of the $SO(2)$ factor of the maximal compact subgroup of the
AdS isometry group. Partially massless
representations are obtained by searching for null descendants. We
thank M. Vassiliev for pointing this out to us.}.  Secondly, we will use the 
above systematics to reassess the case for (and against) extending models of 
cosmological supergravity from their natural AdS arena to~dS.  Before 
proceeding to details, we immediately disclaim any global ambitions for~dS:  
we will be working primarily on one patch of the full~dS, and 
there only within its intrinsic event horizon, this being the lamppost under 
which the physics is (relatively) clear.

In Section~\ref{dS_space}, 
we review the description of~dS as a coset and the corresponding 
algebra as that of the Euclidean conformal group in one lower
dimension.  
Section~\ref{SCAL} 
illustrates our method in the simplest case of scalar fields.  We will 
relate fields in $\rm dS_{n+1}$ 
with conformal ones on ${\mathbb R}^n$ by using an intertwiner, 
essentially establishing a (dS/CFT) correspondence in terms of a proper time 
propagator.  In Section~\ref{BOSONS}, 
we generalize to bosons of arbitrary spin.  We next 
apply the above results to partially massless higher spin bosons, where 
discrete ratios of (mass)$^2$ to cosmological constant imply light cone 
propagation and (partial) gauge invariances.  In
Section~\ref{Fermions} we treat fermions in  
the {\it a priori} hostile~dS context.  Finally, in Section~\ref{SUGRA}, we
revisit the old  
question of the extent, if any, to which a~dS supergravity can be viable, 
presenting the pros and cons in terms of our framework.

\section{de Sitter Space}

\label{dS_space}

The~dS$_{n+1}$ spacetime is the coset $SO(n+1,1)/SO(n,1)$
and is geometrically described by 
the one sheeted hyperboloid\footnote{Our index conventions
are $M,N=0,\ldots {n+1}$, raised and lowered by the Minkowski metric of the
$(n+1)+1$ dimensional embedding space 
$\eta_{MN}={\rm diag}(-,+,\ldots,+)_{MN}$;   
$i,j=1,\ldots,n$ (also denoted by vectors),
covariant with respect to the $SO(n)$ subgroup generated by $M_{ij}$, raised
and lowered by the Kronecker delta. 
}
\be
Z^MZ_M=-(Z^0)^2+\sum_{i=1}^n(Z^i)^2+(Z^{n+1})^2=1\, ,
\label{hyperboloid}
\ee 
embedded in $\mathbb R^{({n+1},1)}$. The cosmological constant
$\Lambda$ 
is positive in~dS space and usually enters the
equation~\eqn{hyperboloid} 
as $n/\Lambda$ on the right hand side. We will work in units $\Lambda=n$ 
throughout, 
however.

The~dS group $SO({n+1},1)$
acts naturally on $\mathbb R^{({n+1},1)}$ with generators
\be
M_{MN}=i(Z_M\d_N-Z_N\d_M)
\ee
obeying the Lie algebra
\be
[M_{MN},M_{RS}]=i\eta_{NR}M_{MS}-i\eta_{NS}M_{MR}
+i\eta_{MS}M_{NR}-i\eta_{MR}M_{NS}\, .
\label{Lie}
\ee
We will employ a coordinate system 
in which spatial sections are flat and therefore rewrite this algebra
in terms of
\be
P_i\equiv M_{i0}+M_{{n+1},i}\, ,\qquad
D\equiv M_{{n+1},0}\, ,\qquad
K_i\equiv M_{i0}-M_{{n+1},i}\, ,
\label{gens0}
\ee
and $M_{ij}$ so that
\bea
&[P_i,P_j]=0\, ,\qquad [K_i,K_j]=0\, ,&\nn\\ \nn\\
&{}[D,P_i]=iP_i\, ,\qquad [D,M_{ij}]=0\, ,\qquad [D,K_i]=-iK_i\, ,&\nn\\ \nn \\
&{}[M_{ij},P_k]=i(P_i\delta_{jk}-P_j\delta_{ik})\, ,\qquad
{}[M_{ij},K_k]=i(K_i\delta_{jk}-K_j\delta_{ik})\, ,&\nn\\\nn\\
&{}[P_i,K_j]=-2i\delta_{ij}D+2iM_{ij}\, .&\label{lie}
\eea
The generators 
$M_{ij}$ satisfy the $so(n)$ angular momentum Lie algebra, so identifying 
the remaining generators $P_i$ as momenta, $D$ as dilations
and $K_i$ as conformal boosts, we see that the algebra~\eqn{lie} generates
the Euclidean conformal group in $n$ dimensions.
To make the relation between the coset $SO({n+1},1)/SO(n,1)$ and
hyperboloid~\eqn{hyperboloid}
explicit, we begin by introducing the
matrix representation of the generators~\eqn{gens0}
\be
P_i=i
\left(
\begin{array}{c|c|c}
0&\vec e_i&0\\\hline
\vec e_i^\top&\!\!\!\!{}_{\phantom{A^A}}0^{\phantom{A^A}}\!\!\!\!&
\vec e_i^\top\\\hline
0&-\vec e_i&0
\end{array}
\right) ,\quad\!\!\!\!
D=i
\left(
\begin{array}{c|c|c}
0&0&-1\\\hline
0&0&0\\\hline
-1&0&0
\end{array}
\right) ,\quad\!\!\!\!
K_i=i
\left(
\begin{array}{c|c|c}
0&\vec e_i&0\\\hline
\vec e_i^\top&
\!\!\!\!{}_{\phantom{A^A}}0^{\phantom{A^A}}\!\!\!\!&-\vec e_i^\top\\\hline
0&\vec e_i&0
\end{array}
\right) .
\label{gens}
\ee
[All entries of the $n$ dimensional vector $\vec e_i$ 
vanish save for the $i^{\rm th}$ slot which is unity.]
Now choose a standard vector $X^M=(0,\ldots,0,1)$. Since this is the
defining representation of $SO({n+1},1)$, the vector $Z^M=(gX)^M$
satisfies the hyperboloid condition~\eqn{hyperboloid} for any $g\in
SO(n+1,1)$ and in fact the entire~dS space can be obtained this
way. The stabilizer of $X^M$ is the subgroup $SO(n,1)\subset SO(n+1,1)$,
therefore $dS_{n+1}\cong SO(n+1,1)/ SO(n,1)$.

\vspace{.3cm}
\noindent
Parameterizing coset elements as
\be
g=\exp(-i\vec P\cdot \vec x)\exp(iDt)
=\left(
\begin{array}{c|c|c}
1+\frac12\Vec x^2&\vec x&\frac12\Vec x^2\\\hline
\;\; \vec x^{\top\phantom{^t}}&I&\; \vec x^\top\\\hline
-\frac12\Vec x^2\;\; &-\vec x\;\;&1-\frac12\Vec x^2
\end{array}
\right)\!
\left(
\begin{array}{c|c|c}
\ch t&0&\sh t\\\hline
0&I&0\\\hline
\sh t&0&\ch t
\end{array}
\right)
,
\label{param}
\ee
we find the embedding coordinates to be 
\be
Z^M=(gX)^M=
\Big(
\sh t+\frac12\, e^t \Vec x^2\, ,\,  e^t\vec x\, ,
\,  \ch t - \frac12\, e^t \Vec x^2
\Big)\, .
\label{embed}
\ee
The~dS metric then follows:
\be
ds^2=dZ^M \eta_{MN} dZ^M=-dt^2+e^{2t}d\Vec x^2
\equiv dx^\mu g_{\m\n}dx^\n\, .
\label{metric}
\ee
While the parameterization of the
coset~\eqn{param} only spans one half, $Z^{n+1}>Z^0$, of the hyperboloid,
the physical region within the intrinsic horizon
\be
0>-1+e^{2t} \Vec x^2=\xi^\mu \xi_\mu\, ,\qquad \xi^\mu=(-1,x^i)\, ,
\label{horizon}
\ee 
is covered by this coordinate patch.

\section{Representing Fields in de Sitter: Scalars}

\label{SCAL}

We now use the simplest, scalar, fields in $\mbox{dS}_{n+1}$ to
illustrate the correspondence with conformal fields on $\mathbb R^n$
based on an intertwiner for on-shell~dS field representations\footnote{
See also~\cite{Dobrev:1998md} for a discussion of the AdS/CFT correspondence
in terms of free field intertwiners.}.

A scalar field in a gravitational background is described by the action
\be
S=\frac12\int d^{n+1}x\sqrt{-g}\, \varphi\Big(D_\mu D^\mu
-\frac{n-1}{4n}\, R
-m^2\Big)\varphi\, ,
\ee
and yields a conformally improved scalar at $m^2=0$.
For the constant curvature~dS background
with metric~\eqn{metric}, the scalar curvature $R=n(n+1)$.
The action is then
invariant under the isometries of $\mbox{dS}_{n+1}$ (for any value of
the mass $m$)
whose action on the
coordinates $x^\mu=(t,x^i)$ may be deduced by examining 
the transformation of the
embedding coordinates $Z^M$ 
in~\eqn{embed} generated by the matrices~\eqn{gens}.
This yields symmetry transformations of $\varphi(x)$ generated by
\bea
&iP_i\;\ =&\d_i\, ,\label{P}\\
&iD\;\ =&-\frac{d}{dt}+\vec x\cdot\vec \d\, ,\\
&iM_{ij}=&x_i\d_{j}-x_j\d_i\, ,\\
&iK_i\;\ =&
-2x_i\Big(-\frac{d}{dt}+\vec x\cdot\vec \d\Big)+\Big(-e^{-2t}+\vec
x^2\Big)\d_i\, . \label{K}
\eea
We can also read off the $\frac12 (n+1)(n+2)$ Killing vectors of
$\mbox{dS}_{n+1}$ from the Lie derivatives above. In particular
$iD=\xi^\mu \d_\mu$, where $\xi^\mu=(-1,x^i)$ is the timelike Killing
vector within the horizon. Observe that the horizon
condition~\eqn{horizon} is invariant under the action of $D$.

The representation~(\ref{P}-\ref{K}) acting on off-shell fields
$\varphi(x)$
is not reducible, as evidenced by the quadratic 
Casimir
\bea
{\cal C}_2&\equiv&\frac{1}{2}M_{MN}M^{NM}\nn\\
&=&D^2+\frac{1}{2}\Big(P_iK_i+K_iP_i\Big)-\frac{1}{2}M_{ij}M^{ij}\nn\\
&=&-\frac{d^2}{dt^2}-n\frac{d}{dt}+e^{-2t}\Vec \d^2\, .\label{c2}
\eea
Incidentally, 
the quartic Casimir\footnote{For $\mbox{dS}_4$, the quartic Casimir
is just the square of a (five dimensional) ``Pauli-Lubanski''
vector $W_M=\e_{MNRST}M^{NR}M^{ST}$} vanishes for the scalar
representation,
\be
{\cal
C}_4\equiv -M_{MN}M^{NO}M_{OP}M^{PM}+2\, {\cal C}_2^{\ {}^{\ss 2}}
-(n-2)(n-3)\, {\cal C}_2\;=\;0\, .
\ee
The operator ${\cal C}_2$ in~\eqn{c2} is precisely the Laplace-Beltrami
operator $D_\mu D^\mu$ on $\mbox{dS}_{n+1}$, so we obtain an irreducible
representation by solving the field
equation
\be
\Big(D_\mu D^\mu-\frac14(n^2-1)-m^2\Big)\varphi=0\, .
\ee
To this end, we expand $\varphi$ in spatial Fourier modes
\be
\varphi(x)=\int d^nx \, e^{i\vec k\cdot \vec x} \, a(\vec k,t)+\mbox{c.c.}
, \qquad k\equiv|\vec k|\, ,
\ee
and must now solve
\be
\Big(-\frac{d^2}{dt^2}-n\frac{d}{dt}-e^{-2t}
k^2-\frac14(n^2-1)-m^2\Big)\, 
a(\vec k,t)=0\, .
\label{shell}
\ee
This is Bessel's equation in the ``conformal 
time''\footnote{The metric $ds^2=u^{-2}(-du^2+d\vec x^2)$ 
is then manifestly conformally Minkowskian.}
variable $u=-\exp(-t)$ for the function $u^{-n/2}a(k,t)$. Hence we may
express $a(k,t)$ in terms of a Hankel function
\be
a(\vec k,t)=k^{-\nu}e^{-\frac12{nt}}H_\nu(-ke^{-t})\;
a(\vec k)\equiv f_k(t;\nu)\, a(\vec k)\, ,
\ee
with index
\be
\nu=\sqrt{\frac 14-m^2}\, .
\ee
Since the Hankel function is just a plane wave (multiplied 
by a slowly varying time dependent amplitude) for index
$\nu=1/2$, conformally improved scalars ($m^2=0$)
propagate on the light cone.

On-shell scalar fields are labeled by the arbitrary function $a(\vec
k)$, so we now study the irreducible representation of the~dS
group acting on these functions induced by the original representation
in terms of isometries of $\mbox{dS}_d$. In mathematical terms, we
want to study the intertwiner
\be
\varphi(x)=\int d^n x\,  f_k(t;\nu)\, e^{i\vec
k\cdot \vec x}\, a(\vec k)+\mbox{c.c.}
\ee
between the isometry representation on functions $\varphi(x)$ and 
the irreducible representation on functions
$a(\vec k)$.

The first step is the intertwining spatial fourier transform
which replaces $x^i\rightarrow i\d/\d k_i$ and $\d/\d x^i\rightarrow
ik_i$ (maintaining the ordering). Denoting the resulting generators
obtained from~\eqn{P} to~\eqn{K} this way by tildes, we have the 
useful identities
\be
i\wt D \, f_k(t;\nu)=\Big(\nu-\frac{n}{2}\Big)\, f_k(t;\nu)\, ,\qquad
i\wt K_i \, f_k(t;\nu)=0\, .
\ee
Therefore we get a very simple action on functions $a(\vec k)$
(dropping the irksome tildes)
\bea
iP_i&=&ik_i\, ,\\
iD&=&-\vec k\cdot\frac{\d}{\d\vec k}+\nu-\frac{n}{2}\, ,\\
iM_{ij}&=&k_i\frac{\d}{\d k^j}-k_j\frac{\d}{\d k^i}\, ,\\
iK_i&=&2i\Big(\vec k\cdot\frac{\d}{\d\vec k}+\frac{n}{2}-\nu\Big)
\frac{\d}{\d k^i}
-ik_i\, \frac{\d}{\d\vec k}\cdot\frac{\d}{\d\vec k}\; .
\eea

Performing an inverse spatial fourier transform, $k_i\rightarrow
-i\d/\d y^i$ and $\d/\d k_i\rightarrow-i y^i$ intertwines to 
the usual representation of the Euclidean conformal group
acting on scalars in $n$ dimensions
\bea
iP_i&=&\frac{\d}{\d y^i}\, ,\\
iD&=&\vec y\cdot \frac{\d}{\d \vec y}+\Delta\, ,\\
iM_{ij}&=&y_i\frac{\d}{\d y^j}-y_j\frac{\d}{\d y^i}\, ,\\
iK_i&=&-2y_i\Big(\vec y\cdot \frac{\d}{\d \vec y}+\Delta\Big)
+\Vec y^2\,\frac{\d}{\d y^i}\,  .
\label{conformal}
\eea
This is precisely the action of the conformal group acting on 
scalars (quasi-primaries) of conformal weight 
\be
\Delta=\frac{n}{2}+\sqrt{\frac14-m^2}\, .
\ee
The intertwiner is given by
\be
\varphi(x^\mu)=\int d^nk\,  e^{i\vec k\cdot\vec x}\, 
f_k(t;\nu)\int d^ny\,  e^{-i\vec k\cdot\vec y} \Phi(\vec y)\,
+\mbox{c.c.}
\; .\label{intertwine}
\ee
The field $\Phi(\vec y)$ is a quasi-primary of weight $\Delta$.
The $\vec k$ integral can be performed by employing the
integral representation of the Hankel function
\be
k^{-\nu}H_\nu(uk)=-\frac{i}{\pi}\, e^{-\frac12i\nu\pi}\, 
\int_0^\infty \frac{d\tau}{\tau^{\nu+1}}\, 
\exp\Big[\frac{1}{2}\, iu\Big(\tau+\frac{k^2}\tau\Big)\Big]\, .
\ee
We find 
\be
\varphi(x^\mu)= \int d^ny\, \Delta(\vec x-\vec y,t)\, 
\Phi(\vec y)\, .
\ee
Here
\be
\Delta(\vec x-\vec y,t)={\cal N}
\int_0^\infty
\frac{d\tau}{\tau^{\nu-\frac{n}{2}+1}}\exp
\Big[-\frac{i\tau}{2u}\Big(-u^2+(\vec x-\vec y)^2\Big)\Big]\, .
\ee 
is the Schwinger proper time parametrization of the bulk
to boundary propagator from
conformal time $u=0$ to $u=-\exp(-t)$ (up to some overall
normalization ${\cal N})$. Therefore the
intertwiner~\eqn{intertwine} exactly realizes the~dS/CFT correspondence
of~\cite{Strominger:2001pn}.

We end this section by noting that the quadratic Casimir for the 
conformal representation~\eqn{conformal} is
\be
{\cal C}_2=-\Delta(\Delta-n)=\frac14(n^2-1)+m^2\, ,
\ee
which agrees with the one imposed by the field equation~\eqn{shell}.
The quartic Casimir invariant vanishes also in the 
conformal representation, as it should for scalars.

\section{Representing Arbitrary Spin Bosons}

\label{BOSONS}

A massive bosonic spin $s$ field in~dS space can be
described by a completely symmetric\footnote{We concentrate on 
 completely symmetric higher spin representations. 
See~\cite{Brink:2000ag,Zinoviev:2002ye} 
for a discussion of higher spins for other tensor symmetries.
It is also worth noting that to obtain the simple field equation
and constraints in~\eqn{eoms} from an action for spins $s\geq 5/2$,
it is necessary to introduce auxiliary fields. They play, however, no
r\^ole in the representation theoretic analysis given here.} 
tensor $\varphi_{\mu_1\ldots\mu_s}$
subject to the field equation and constraints
\be
\Big(D_\mu D^\mu-2n+4+(n-5)s+s^2-m^2\Big)\varphi_{\mu_1\ldots\mu_s}=0=
D.\varphi_{\mu_2\ldots\mu_s}=\varphi^\rho{}_{\rho\mu_3\ldots\mu_s}\, .
\label{eoms}
\ee
For generic values of the mass $m$, $\varphi_{\mu_1\ldots\mu_s}$ describes the
\be
\mu(n,s)\equiv
\frac{(n+2s-2)(n+s-3)!}{s!(n-2)!}
\ee
degrees of freedom of a spin $s$ symmetric field in $n+1$ dimensions. 
The mass parameter\footnote{For scalars, there is no
gauge invariance, but one often chooses $m^2$ such that vanishing 
mass yields a conformally improved scalar in general backgrounds as we did
in Section~\ref{SCAL}. It is a peculiarity of four dimensions that
the choice of mass parameter required for gauge invariance of spins $s\geq1$
also yields the conformally improved scalar when continued to $s=0$.} 
is defined so that the theory is strictly massless
for $m^2=0$ with a gauge invariance\footnote{For example, the massive spin~1
field equation ${\cal G}_\mu=(D^2-n-m^2)\varphi_\mu
-D_\mu D.\varphi$ obeys a Bianchi identity $D.{\cal G}=0$ at $m^2=0$,
as is easily verified using
$[D_\m,D_\n]\varphi_\rho=2g_{\rho[\m}\varphi_{\nu]}$ in~dS.}
\be\varphi_{\mu_1\ldots\mu_s}=\d_{(\mu_1}\xi_{\mu_2\ldots \mu_s)}\, ,
\label{gauge}\ee (subject
to $\xi^\rho{}_{\rho\mu_3\ldots\mu_s}=0$).
The degree of freedom count is then
\be
\mu(n,s)-\mu(n,s-1)=\frac{(n+2s-3)(n+s-4)!}{s!(n-3)!}\, .
\label{countfreedom}
\ee
Actions may be written down
for these free theories, both massive and massless; 
see~\cite{Deser:2001us,Zinoviev:2001dt} for details.
The~dS isometries act on off-shell fields $\varphi_{\mu_1\ldots\mu_s}$
as
\bea
iP_i \varphi_{\mu_1\ldots\mu_s}&=&\d_i\varphi_{\mu_1\ldots\mu_s}\, ,\\
iD \varphi_{\mu_1\ldots\mu_s}&=&\Big(-\frac{d}{dt}+\vec x\cdot\vec \d\Big)\
\varphi_{\mu_1\ldots\mu_s}+s\, \delta_{(\mu_1}^j
\varphi_{\mu_2\ldots\mu_s)j}\, ,\\
iM_{ij}\varphi_{\mu_1\ldots\mu_s}&=&(x_i\d_j-x_j\d_i)\varphi_{\mu_1\ldots\mu_s}
-s\, \varphi_{i(\mu_1\ldots}\delta_{\mu_s)j}+
s\, \varphi_{j(\mu_1\ldots}\delta_{\mu_s)i}\, ,\\
iK_i\varphi_{\mu_1\ldots\mu_s}&=&
\Big[-2x_i\Big(-\frac{d}{dt}+\vec x\cdot\vec \d\Big)+\Big(-e^{-2t}+\vec
x^2\Big)\d_i\Big]\, \varphi_{\mu_1\ldots\mu_s}\nn\\
&+&2s\, \delta_{i(\mu_1}\varphi_{\mu_2\ldots\mu_s)0}
+2s\,e^{-2t} \delta^0_{(\mu_1}\varphi_{\mu_2\ldots\mu_s)i}
-2s\, \delta_{i(\mu_1}\varphi_{\mu_2\ldots\mu_s)j}x_j\nn\\
&+&s\, x_j\Big(\varphi_{i(\mu_1\ldots}\delta_{\mu_s)j}-
\varphi_{j(\mu_1\ldots}\delta_{\mu_s)i}\Big)\, .
\eea

It is tedious but not difficult 
to solve the field equations in the frame~\eqn{metric} and then
construct the higher spin bulk to boundary propagator
\be
\varphi_{\mu_1\ldots\mu_s}=\int d^ny \Delta(\vec x-\vec
y,t)_{\mu_1\ldots\mu_s}
^{i_1\ldots i_s} V_{i_1\ldots i_s}\, .
\ee
The higher spin boundary fields $V_{i_1\ldots i_s}$ are totally symmetric and 
traceless in $n$ dimensions and transform under the 
conformal algebra as
\bea
iP_i V_{(s)}&=&\d_i V_{(s)}\, ,\\
iD_i V_{(s)}&=&(y^i \d_i+\Delta_s) V_{(s)}\, ,\\
iM_{ij}V_{(s)}&=&(y_i \d_j-y_j\d_i)-{\cal S}_{ij} V_{(s)}\, ,\\
iK_iV_{(s)}&=&(-2iy_iD+i\Vec y^2 P_i)V_{(s)}+y_j{\cal S}_{ij}V_{(s)}\, .
\eea
Here $V_{(s)}$ is shorthand for $V_{i_1\ldots i_s}$
and the intrinsic spin operator ${\cal S}_{ij}$
acts as
\be
S_{ij} V_{i_1\ldots i_s}=\sum_{\alpha=1}^s 
V_{i_1\ldots i_{\alpha-1}ii_{\alpha+1}\ldots i_s}\delta_{ji_\alpha}-
(i\leftrightarrow j)
\, .
\ee
The weight $\Delta_s$ is computed by examining the quadratic Casimir
of this representation
\be
{\cal C}_2=-\Delta_s(\Delta_s-n)-s(s+n-2)\, .
\label{C2}
\ee
For higher spins, the quadratic Casimir and Laplacian are no longer
equal; instead, a simple computation reveals 
(see~\cite{Pilch:1984xx,deWit:2002vz}) that
\be
D_\mu D^\mu\varphi_{\mu_1\ldots\mu_s}=\Big({\cal C}_2+s(s+n-1)\Big)
\varphi_{\mu_1\ldots\mu_s}\, . 
\label{shift}
\ee
Comparing~\eqn{C2},{}~\eqn{shift} and~\eqn{eoms} relates the
mass parameter $m^2$ and weight $\Delta_s$,
\be
m^2=-\Delta_s(\Delta_s-n)+(s-2)(s-2+n)\, .\label{mass}
\ee
Our next task is to find the values of $m^2$ making fields
massless or partially massless.

\section{Partially Massless Bosons}

We have assembled all the relevant machinery to provide a very
simple description of partially massless bosons in 
dS$_{n+1}$ in terms of representations of the $n$ dimensional
conformal group\footnote{These representations have been studied
in a conformal field theory context in~\cite{Dobrev:qv,Dobrev:1976ru,
Dobrev:1978fv,Todorov:rf}.}.

We recall that the physical polarizations of a 
strictly massless field satisfy 
\be
\d.V_{(s-1)}\equiv \d^i V_{ii_2\ldots i_s}=0\, .
\label{strict}
\ee
thanks to the gauge invariance~\eqn{gauge} which projects out
all but the maximal helicity $s$ excitations\footnote{Helicity in
dimensions $n>3$ can be defined in terms of $\vec P^{-2}(\e^{i_1\ldots
i_n}P_{i_1}M_{i_2i_3})^2$. Strictly speaking, in the discussion above,
we should refer to the absolute value of the helicity.}. 
A field obeying~\eqn{strict} has the correct degree of freedom 
count, as given in~\eqn{countfreedom}, for a strictly massless field.

For partially massless fields,  gauge invariances of the form
\be
\delta \varphi_{\mu_1\ldots\mu_s}=\D_{(\mu_1\ldots\mu_t}
\varphi_{\mu_{t+1}\ldots\mu_s)}+\cdots
\ee 
imply that the requirement~\eqn{strict} is
relaxed and replaced by
\be
\d^{i_1}\cdots \d^{i_t} V_{i_1\ldots i_s}=0\, , \qquad (t\leq s).
\label{partial}
\ee
We call such a field ``partially massless of depth $t$''.
This amounts to projecting out all helicities save $(s,\ldots,t+1)$
and gives $\mu(n,s)-\mu(n,s-t)$ degrees of freedom.

The subalgebra of translations, dilations and rotations 
leaves the condition~\eqn{partial} invariant. However, conformal
boosts do not, unless one tunes the conformal weights $\Delta_s$
appropriately. To obtain these tunings we study
\be
\d^{i_1}\cdots\d^{i_t}K_iV_{i_1\ldots i_s}=0\, ,
\label{condition}
\ee
for depth $t$ partially massless polarizations 
$V_{(s)}$ subject to~\eqn{partial}.
It is a simple combinatorics problem to compute the (unique)
value of $\Delta_s$ as a function of the depth $t$ such that 
the condition~\eqn{condition} holds. We state the result below.
The main idea is conveyed by the simplest non-trivial example, spin 2.

For a spin 2 field $V_{ij}-V_{ji}=0=V_i{}^i$, the conformal boost
acts as
\be
iK_iV_{jk}=i(-2y_iD+\Vec y^2 P_i)V_{jk}+4y_{(j}V_{k)i}
-\delta_{i(j}y_lV_{k)l}\, .
\ee
The field $V_{ij}$ is strictly massless whenever
\be
\d^iV_{ij}=0\, , \label{div}
\ee
so we test whether this condition is respected by conformal
boosts by computing
\be
\d^k\, K_iV_{jk}=2i(\Delta_s-n)V_{ij}\, .
\ee
Here we have relied on the divergence constraint ~\eqn{div}. 
Hence we find the strictly
massless tuning 
\be
\Delta_s=n\, .\label{stricttune}
\ee
Using this relation as well as $s=2$ in~\eqn{mass} gives $m^2=0$, the
correct tuning for a strictly massless spin~2 boson.

To study partially massless spin~2, we replace the single
divergence condition~\eqn{div} by the double divergence one
\be
\d^{ij}V_{ij}=0.\label{divdiv}
\ee
Now we must compute
\bea
\d^{j}\d^k\, K_iV_{jk}&=&4i(\Delta_s-n+1)\d_jV_{ij}
\eea
where we used~\eqn{divdiv} but {\it not}~\eqn{div}.
Therefore we obtain the partially massless spin~2 
tuning
\be
\Delta_s=n-1\, .\label{partune}
\ee
It is also clear that the partially massless representation
is irreducible. One might have thought it to be a direct sum of
spin~2 and spin~1 strictly massless representations. However, since
the tunings~\eqn{stricttune} and~\eqn{partune} differ, 
the strictly massless spin~2 field components satisfying~\eqn{div}
mix with the remaining ones when $\Delta_s\neq n$.

We calculate the tunings for arbitrary spin in the same way
by imposing~\eqn{partial}
and computing
\be
\d^{i_1}\cdots\d^{i_t} K_i V_{i_1\ldots i_s}=
2it(\Delta_s-n-s+t+1)\, \d^{i_2}\cdots\d^{i_{t}}V_{ii_2\ldots i_s}\, .
\ee
The tunings are therefore
\be
\Delta_s=n+s-t-1\, .\label{loony_tunes}
\ee
Inserting the tuning condition~\eqn{loony_tunes} in
the~dS mass--conformal weight relation~\eqn{mass}
yields the mass conditions for depth $t$ partial masslessness
\be
m^2=(t-1)(2s-3+n-t)\, .
\ee
Firstly note that for depth $t=1$, {\it i.e.} strictly massless
fields, the mass parameter $m^2=0$. Also when $n=3$, the result agrees
with the one conjectured in~\cite{Deser:2001xr} on the basis of requiring
light cone propagation for partially massless fields. 

\section{Fermion Representations}

\label{Fermions}

There is no fundamental difficulty in adapting the above bosonic
manipulations to fermions, only tedium. 
Rather than performing
the computation we note that a massive spin $s\equiv \sigma+1/2$
fermionic field satisfies 
\be
(\slash\!\!\!\!
D+m)\psi_{\mu_1\ldots\mu_s}=0=D.\psi_{\mu_2\ldots\mu_s}=
\gamma.\psi_{\mu_2\ldots\mu_s}\, .
\ee
Here $\psi_{\mu_1\dots\mu_s}$ is a completely symmetric tensor-spinor.
With the above choice of the parameter $m$, strict masslessness occurs
at\footnote{Again, this criterion applies to masslessness imposed by a
gauge invariance and is valid for spins $s\geq3/2$. In four dimensions,
coincidentally, it yields $m^2=0$ for $s=1/2$, the same value required for a
conformally improved spinor.} 
\be
m^2=-\Big(s+\frac n2-2\Big)^2\, .
\ee
Accounting for this ``offset'', we find partially massless tunings
for fermions at
\be
m^2=-\Big(s-t+\frac n2-1\Big)^2\, ,
\label{ftune}
\ee
in precise agreement with the result of~\cite{Deser:2001xr} for $\mbox{dS}_4$.
Recalling that the cosmological constant is reinstated by
multiplying the right hand side of~\eqn{ftune} by $\Lambda/n$, we note
(as observed in~\cite{Deser:2001pe}), that fermionic tunings are satisfied for 
real values of $m$ only for $\Lambda<0$, {\it i.e.} in AdS space.
The parameter $m$, (not its square) appears in the action, for these
theories. For the choice of square root 
$m=+\, i\ (s-t+n/2-1)$, the action 
for partially (and strictly) massless~dS fermionic fields no longer 
obeys a reality condition, but is invariant under a formal gauge invariance.
This property is the root cause of the difficulty defining~dS
supergravities, the topic of the next section.  

\section{de Sitter Supergravity Revisited}
\label{SUGRA}
Our considerations thus far have led us to the following picture 
(illustrated in Figure~\ref{linepik}) of
particles in (A)dS backgrounds. Partially massless fields, of either 
statistics, are always unitary in~dS, while in AdS only strictly massless ones 
are. From Figure~\ref{linepik}, 
this behavior is understood by turning on cosmological 
constants of either sign and following their effects on the signs of lower 
helicity state norms. A sequence of unitary partially massless
fields is only encountered when starting from Minkowski space
($\Lambda=0$) and first turning on a positive cosmological constant.
The bad news, 
however, is that partially massless
fermions require tunings with negative $m^2$ as already noted in cosmological  
supergravity~\cite{Townsend:1977qa,Freedman:1977aw}.   
This led to the rejection of (see especially~\cite{Pilch:1985aw}) 
dS supergravity   
as a consistent local QFT, a rejection bolstered later by the  difficulties   
in defining 
string theory on~dS backgrounds (see for example~\cite{Witten:2001kn}).

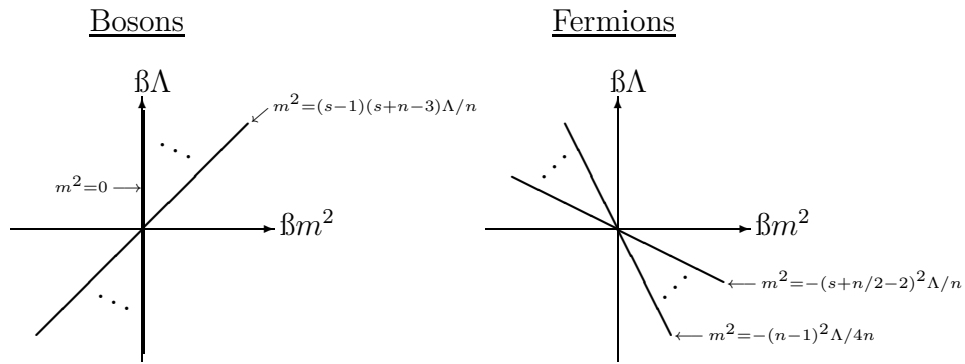
\begin{figure}
\begin{picture}(100,300)(-20,0)

\put(30,125){\underline{Bosons}}
\put(205,125){\underline{Fermions}}

\put(102,48){$\ss m^2$}
\put(47,102){$\ss \Lambda$}
\put(17,64){$\sss m^2=0\, \longrightarrow$}
\put(91,95){${}_{_{\swarrow}}\sss m^2=(s-1)(s+n-3)\Lambda/n$}
\put(55,80){\rotatebox{-25}{$\cdots$}}
\put(32,22){\rotatebox{155}{$\cdots$}}

\put(282,48){$\ss m^2$}
\put(227,102){$\ss \Lambda$}
\put(271,28){{$\sss \longleftarrow\, m^2=-(s+n/2-2)^2\Lambda/n$}}
\put(251,8){{$\sss \longleftarrow\, m^2=-(n-1)^2\Lambda/4n$}}

\put(245,21){\rotatebox{45}{$\cdots$}}
\put(199,81){\rotatebox{225}{$\cdots$}}

\put(0,50){\vector(1,0){100}}
\put(50,0){\vector(0,1){100}}

\put(180,50){\vector(1,0){100}}
\put(230,0){\vector(0,1){100}}

\thicklines

\put(10,10){\line(1,1){80}}
\put(50.5,3){\line(0,1){92}}

\put(270,30){\line(-2,1){80}}
\put(250,10){\line(-1,2){40}}
\end{picture}

\caption{\label{linepik} Cosmological phase diagrams for partially
massless Bose and Fermi fields depicting strictly massless
and maximal depth partially massless tunings. All other lines
with partially massless gauge invariances appear between these.}

\end{figure}

Let us first present the reasons for rejection in terms of the present 
analysis, followed by such mitigating circumstances as we can muster for 
keeping the possibility in play. For concreteness, we deal primarily here
with $N=1$ supergravity in four dimensions.

\begin{itemize}
\item Although (formally) locally supersymmetric actions exist, the mass
parameter appearing in the term 
$m\sqrt{-g}\, \psi_\m\gamma^{\m\n}\psi_\n$ must be pure imaginary for 
lightcone propagation.
Therefore, the action of~dS supergravity does not obey
a reality condition. 

\item The associated local supersymmetry
transformations $\delta\psi_\mu=(D_\mu+\frac12 m\gamma_\mu)\varepsilon$,
being complex, cannot
preserve Majorana
condition on fields. For $N=2$ supersymmetry, a symplectic
reality condition is possible, but the locally supersymmetric action is
either still complex or the Maxwell field's kinetic term has
tachyonic sign~\cite{Pilch:1985aw}.

\item Another way to see that there can be
no real supercharges is that the $N=1$~dS superalgebra 
ought be the $d=5$, $N=1$ superlorentz algebra, but
there are no Majorana spinors in $d=5$ Minkowski space. 

\item Because~dS$_4$ has topology $S^3\times
{\mathbb R}$, gauge charges (being surface integrals) 
vanish since there are no spatial boundaries~\cite{Witten:2001kn}. 
This argument is of a different nature from the previous ones
as it involves the global considerations which we have chosen to ignore.

\item Finally, even for the $N=2$ case, where a~dS superalgebra
exists, there are no positive mass, unitary representations~\cite{Pilch:1985aw}. 
Unlike the praiseworthy AdS algebra, our maximal compact subalgebra
has no $SO(2)$ factor whose generator could define a positive mass
Hamiltonian. Again this is a global issue.
\end{itemize}

Let us now present the arguments in favor of~dS supergravity:
\begin{itemize}
\item While the tuned~dS supergravity mass
\be
m^2=-\Lambda/3\, .
\ee 
is indeed negative, there 
are precedents for consistent theories with negative~$m^2$, for example
scalars in AdS for which a range of such values can be 
tolerated~\cite{Breitenlohner:1982bm,Breitenlohner:1982jf}
essentially because the lowest
eigenvalue of the Laplacian has a positive offset there.
Figure~\ref{linepik} shows that fermions mirror this behavior in~dS.

\item Despite an imaginary
mass-like term in the action, at least the free field
representations are unitary. For unitarity of spin~3/2
representations, the
relevant quantity is $m^2-3\Lambda$, not $m^2$ alone. In addition the
linearized equations of motion for physical spin~2, helicity~$\pm2$ and
its proposed spin~$3/2$, helicity~$\pm3/2$ superpartner degrees of freedom
are identical
\bea
\mbox{Spin~2:}\qquad
\Big(-u^2\frac{d^2}{du^2}-u\frac{d}{du} 
+u^2\, \Vec \d^2+\frac{3}{4}\, \sqrt{\Lambda}
\Big)\, \epsilon_{\pm 2}=0\, ,\\ 
\mbox{Spin~3/2:}\quad
\Big(-u^2\frac{d^2}{du^2}-u\frac{d}{du} 
+u^2\, \Vec \d^2+\frac{3}{4}\, \sqrt{\Lambda}
\Big)\, \epsilon_{\pm 3/2}=0\, .
\eea
Here we have employed
the frame $ds^2=u^{-2}(-du^2+d\Vec x^2)$; a detailed derivation
may be found in~\cite{Deser:2001xr}.

\item In~dS space positivity of energy is possible only for
localized excitations within the horizon~\eqn{horizon}. Only this
region of~dS possesses a timelike Killing vector.~dS
gravity is therefore stable against local excitations within the
physical region~\cite{Abbott:1982ff}\footnote{The same conclusion applies
also to partially massless~dS fields, see~\cite{Deser:2001wx}.}. 
Observations of supernovae
suggest that we inhabit an 
asymptotically~dS universe~\cite{Carroll:2000fy}. Although~dS
quantum gravity is problematic, nobody would reject cosmological
Einstein gravity as an effective description of local physics
within the physical region. This looser criterion 
is all we should require of a
sensible~dS supergravity.
\item
As stated, any local supersymmetry is purely formal in the absence
of $d=4+1$ Majorana spinors (even the fermionic field equation above 
is not consistent with a reality condition on $\psi_\mu$).
In any case, a bona fide $N=1$~dS superalgebra with
Majorana super charge, would imply a global positive energy theorem from 
$\{Q,Q\}\sim H$, which is already ruled out at the level of~dS
gravity. Instead, we could envisage relaxing the Majorana
condition
and allowing only a formal local supersymmetry.
A direct Hamiltonian constraint analysis still
shows that only helicities~$\pm3/2$
propagate. Also there are other examples where the Majorana condition
must be dropped, but nonetheless a formal supersymmetry yields the
desired Ward identities: continuation to Euclidean space is a familiar
case~\cite{Nicolai:1978vc,Nicolai:1979ic,Nicolai:1979se,
Nicolai:1980nr,vanNieuwenhuizen:1996tv,Waldron:1998re}.
\end{itemize}

The summary in favor of~dS supergravity is then that 
the free field limit is unitary with
time evolution governed by the Hermitean generator $D=iM_{40}$
subject to a positive energy condition--within the physical region
inside the Killing horizon. A formal supersymmetry, similar to 
that remaining when Euclideanizing supersymmetric Minkowski models,
suffices to show that amplitudes obey supersymmetric Ward identities.

\section{Conclusions}

Our work has consisted of two parts, one unambiguously correct, 
the other more speculative, or better, open-ended.  The first, 
demonstrable, one was devoted to a simple formulation of generic spin massive 
models in arbitrary dimensional~dS, one that was particularly relevant to 
hierarchies of partial massive higher spins.  Use of a~dS/CFT correspondence 
between the fields and their (boundary) Euclidean limits in one lower 
dimension was an important ingredient in the process, and our conclusions 
established and generalized to arbitrary dimensions our original conjectures 
in this respect. In particular we have provided an explicit realization
of the suggestion of~\cite{Dolan:2001ih}, that an AdS/CFT correspondence for 
partially massless fields should be considered in~dS, rather than AdS
where these fields are not unitary. Of course the most pressing question
now is whether this new perspective yields any insight on the much harder
problem of interactions for partially massless fields.

Our second aim was to review the arguments, pro and con, concerning 
existence, albeit in a restrictive sense, of N=1~dS supergravity.  
The arguments against 
are well known, revolving about the need for an imaginary spin~3/2 mass 
parameter and correspondingly imaginary term in its action, all in addition 
to the generic problems inherited from the intrinsic horizon of~dS.  But 
imaginarity in turn implies that the local SUSY is purely formal and the 
supercharges are not Hermitean.  The arguments in favor accept, but try to 
turn, these manifest difficulties into harmless ones.  Whatever their force, 
they do have on their side the fact that we may well be living inside the 
horizon of some asymptotically~dS world, one in which supersymmetric physics 
should have a place.

\section*{Acknowledgements}

It is a pleasure to thank B. Pioline, R. Stanton and 
M. Vasiliev for discussions. 
A.W. thanks Brandeis University and the Max Planck Institut f\"ur Mathematik
Bonn for hospitality. This work was supported in part by NSF grants
PHY99-73935 and PHY01-40365.

\bibliographystyle{h-elsevier2}
\bibliography{dSCFT}

\end{document}